\documentclass[12pt]{iopart}

\usepackage{graphicx}
\begin{document}

\title[From CELSIUS to COSY: on the observation of a dibaryon resonance]{From
  CELSIUS to COSY: on the observation of a dibaryon resonance}

\author{H Clement$^{1,2}$, M Bashkanov$^{1,2}$ and T Skorodko$^{1,2,3}$}

\address{$^{1}$ Physikalisches Institut der Universit\"at T\"ubingen, Auf der
  Morgenstelle 14, D-72076 T\"ubingen, Germany}
\address{$^{2}$ Kepler Center for Astro and Particle Physics, University of
  T\"ubingen, Auf der Morgenstelle 14, D-72076 T\"ubingen, Germany}
\address{$^{3}$ Department of Physics, Tomsk State University, 36 Lenina
  Avenue, 634050 Tomsk, Russia}
\ead{heinz.clement@uni-tuebingen.de}
\vspace{10pt}
\begin{indented}
\item[]October  2014
\end{indented}

\begin{abstract}
Using the high-quality beam of storage rings in combination with a pellet
target and the hermetic WASA detector covering practically the full solid
angle the two-pion production in nucleon-nucleon collisions has been
systematically studied by exclusive and kinematically complete
measurements -- first at CELSIUS and subsequently at COSY. These measurements
resulted in a detailed understanding of the two-pion production mechanism by
$t$-channel meson exchange. The investigation of the ABC effect in
double-pionic fusion reactions lead the trace to the observation of a narrow
dibaryon resonance with $I(J^P) = 0(3^+)$ about 80 MeV below the nominal mass
of the conventional $\Delta\Delta$ system. New neutron-proton
scattering data, taken with polarized beam at COSY, produce a pole in the
coupled $^3D_3$ - $^3G_3$ partial waves at ($2380\pm10~-~i~40\pm5$) MeV
establishing thus the first observation of a genuine $s$-channel dibaryon
resonance.  
\end{abstract}

\pacs{13.75.Cs, 13.85.Dz, 14.20.Pt,25.40.Ve}
%
\vspace{2pc}
\noindent{\it Keywords}: dibaryon, two-pion production
%
%
%
%

\section{Introduction}
\label{intro}
The question whether the two-baryon system possesses more eigenstates than
just the isoscalar deuteron groundstate and the isovector virtual $^1S_0$
state (known from the NN final state interaction) awaits an answer since
decades. This fundamental question has been connected to the question about
six-quark systems first in 1964, when Dyson and Xuong \cite{Dyson}
correlated this topic with symmetry breaking of SU(6) -- just shortly after
Gell-Mann's famous publication  \cite{GellMann} of the quark model. However,
this topic did not find overwhelming attention until 1977, when Jaffe
predicted the so-called H dibaryon \cite{Jaffe}, a bound $\Lambda\Lambda$
system, based on quantum chromodynamics (QCD). It was this paper, which
initiated a real dibaryon rush with a copious number of theoretical
calculations and an even greater number of dibaryon searches resulting in a
vast number of experimental claims --- but finally none survived careful
experimental investigations. For a review see, {\it e.g.}
Ref. \cite{Seth}.

Reasons for this striking failure of previous dibaryon searches may be sought,
at least party, in the insufficient quality of data based either on
low-statistics bubble-chamber or inclusive measurements performed mainly with
single-arm detectors \cite{Seth}. 

Recently the question about dibaryons received renewed interest after it had
been realized that there exist more complex quark configurations than just the
familiar $q\bar{q}$ and $qqq$ systems -- in favor also of hidden color aspects
\cite{BBC}.

\section{Exclusive and kinematically complete experiments -- observation of a dibaryon resonance }
\label{sec-2}

Since two-pion production was hardly investigated, but had the potential to
contain unusual phenomena, a systematic study of this process was started in
the nineties at the CELSIUS storage ring, where by the begin of the new
millennium the WASA detector with close-to $4\pi$ angular coverage 
went into operation in combination with a hydrogen/deuterium  pellet
target \cite{CB}. This provided the possibility to measure meson 
production in $pp$, $pd$ and $dd$ collisions for the first time exclusively and
kinematically completely --  even with up to six overconstraints. In addition
the high-quality beam in combination with the pellet target ensured
a low background situation. That way such measurements delivered data of
unprecedented quality. 

In 2005 the WASA detector was moved to the COSY ring taking advantage of the
still much superior beam quality and intensity there \cite{wasa}. This allowed
to move the studies from $pp$ induced to $pn$ induced reactions by measuring the
quasifree scattering process with high accuracy over a wide energy region.

Additionally, two-pion production was also studied at COSY with polarized beam
utilizing the TOF detector, which  also covered practically the full
solid angle \cite{AE}.

As a result of these kinematically complete and exclusive two-pion
production measurements at CELSIUS and subsequently at COSY it was shown that
all $pp$ induced, {\it i.e.} all isovector channels, behave as
expected from conventional $t$-channel excitation of Roper, $\Delta\Delta$
(mutual excitation of the colliding nucleons into their first excited state
$\Delta(1232)$) and $\Delta(1600)$ resonances
\cite{AE,JJ,WB,JP,TS,FK,deldel,nnpipi,iso,tt}.  

The situation changes strikingly in $pn$ initiated two-pion
production. Especially the purely isoscalar $pn \to d\pi^0\pi^0$ reaction,
for which the cross section from the conventional processes was expected to be
particularly small and for which also no data existed so far, exhibits a
striking narrow Lorentzian structure in the total cross section. First
indications of that were observed at CELSIUS \cite{MB}. Later-on this resonance
structure was measured in detail at COSY with much improved statistics
and accuracy 
\cite{prl2011} leading to m$\approx$ 2370 MeV and $\Gamma \approx$ 70
MeV for mass and width. The latter is more than three times smaller than
expected from a conventional $\Delta\Delta$ excitation via the $t$-channel meson
exchange. Nevertheless the Dalitz plot clearly shows that this resonance
structure decays predominantly via a $\Delta\Delta$ intermediate state,
though the resonance mass is about 80 MeV below the nominal mass
$2m_{\Delta}$ of the $\Delta \Delta$ system. The angular distributions
measured for this reaction determine the spin-parity of this structure to be
$J^P = 3^+$~\cite{prl2011}.

Subsequent measurements of all three reactions leading to the
double-pionic fusion to the deuteron uniquely determined this structure as
being of isoscalar nature. Fig. 1, left, shows the experimental decomposition
of the isospin-mixed $pn \to d\pi^+\pi^-$ reaction (red symbols) into its
isoscalar (blue) and isovector (black) components corresponding to the
$d\pi^0\pi^0$ and $d\pi^+\pi^0$ channels, respectively \cite{isofus}. Note
that the latter does not show the resonance structure. 

If this structure, indeed, represents a genuine $s$-channel resonance, it
has to show up also in the entrance channel, though its effect in  $np$
scattering is expected to be very small \cite{PBC}. For such an investigation
the $np$ analyzing power is the most promising observable. Since it consists
only of partial-wave interference terms, it is particularly sensitive to
small contributions. Hence, in the absence of data from previous experiments,
polarized $\vec{n}p$ scattering 
measurements were conducted with WASA at COSY \cite{prl2014,npfull}. The
resulting data show a resonance structure in the energy dependence -- right at
the expected position. At the right-hand side of Fig. 1, the results are
depicted for a center-of-mass scattering angle near 90$^\circ$, where the
effect of a $J^P = 3^+$ resonance is largest. Inclusion of the WASA data in
the SAID data base and its subsequent partial-wave analysis, indeed, 
produces a pole in the coupled $^3D_3 - ^3G_3$ partial waves at $(2380\pm 10)
- i(40\pm5)$ MeV  -- in agreement with the resonance hypothesis based on the
double-pionic fusion results \cite{prl2014,npfull} and denoted now accordingly
$d^*(2380)$,

\begin{figure}[t]
\centering
\includegraphics[width=5cm,clip]{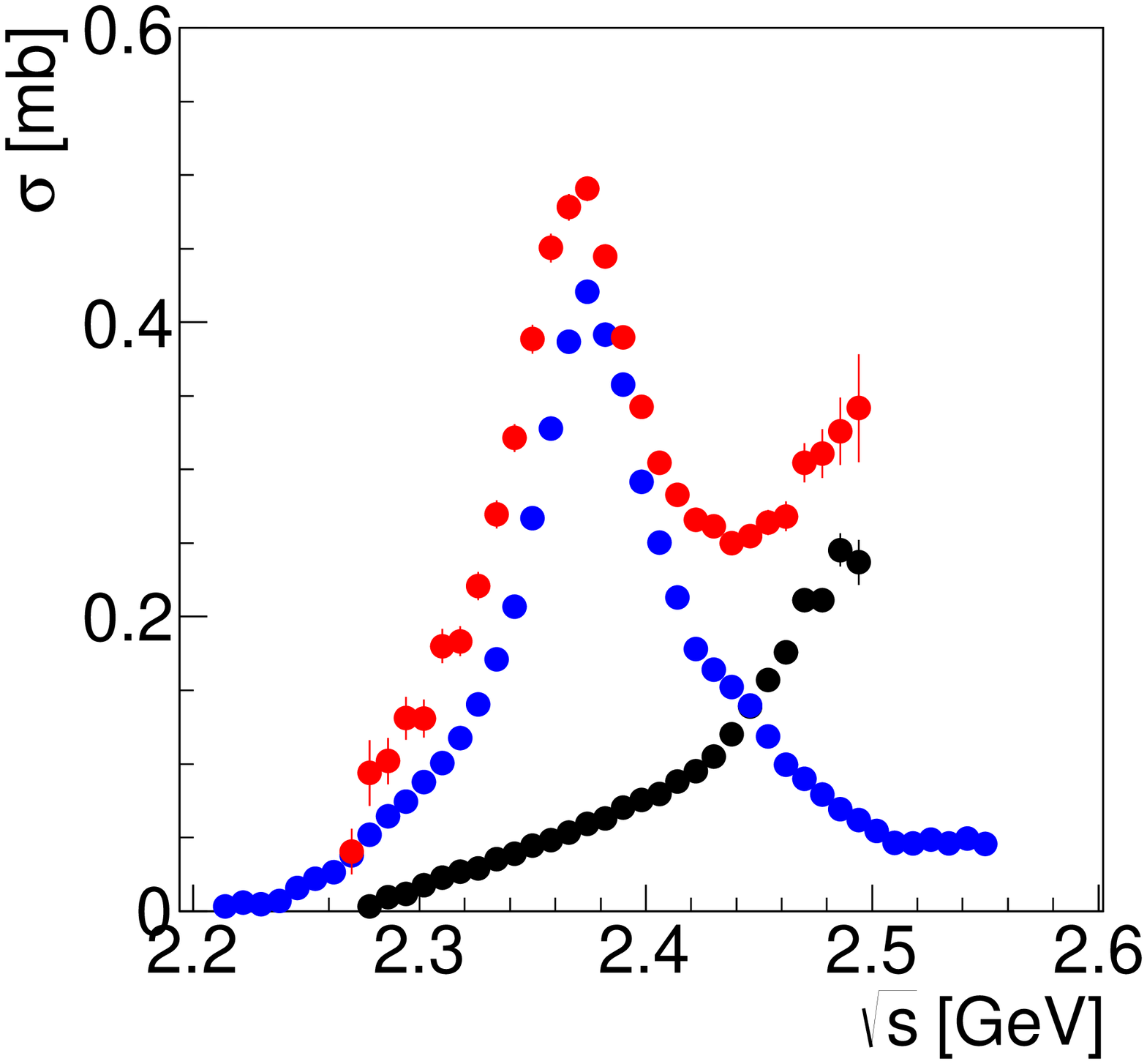}
\includegraphics[width=7cm,clip]{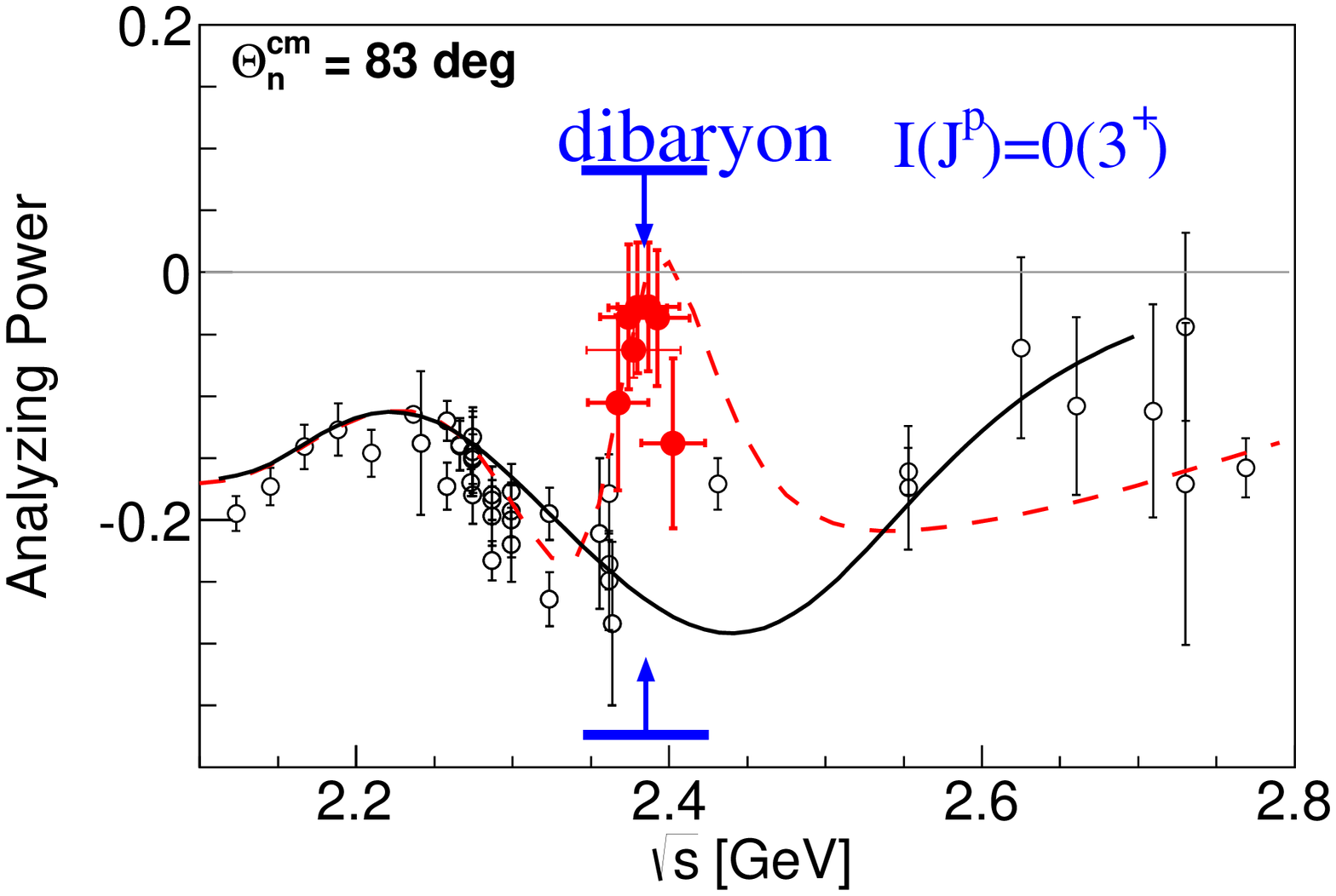}
\caption{Left: Dependence of the total cross section for the
  reaction $pn \to d\pi^+\pi^-$ (red) and its isospin decomposition
  \cite{isofus} into isoscalar part -- corresponding to $2 \sigma(pn \to 
  d\pi^0\pi^0)$ (blue) -- and isovector part -- corresponding to $1/2~\sigma(pp
  \to d\pi^+\pi^0)$ (black) -- on the center-of-mass energy $\sqrt s$. Right:
  Energy  
  dependence of the analyzing power in $\vec{n}p$ scattering near 90$^\circ$,
  where the effect of a $I(J^P) = 0(3^+)$ resonance shows up most
  clearly. Filled circles denote WASA results, open symbols previous
  work. The solid line gives the current SAID solution SP07, the
  dashed line represents the new SAID solution containing the resonance pole
  \cite{prl2014,npfull}.} 
\label{fig-1}       
\end{figure}

\begin{figure} 
\centering
\includegraphics[width=4.9cm,clip]{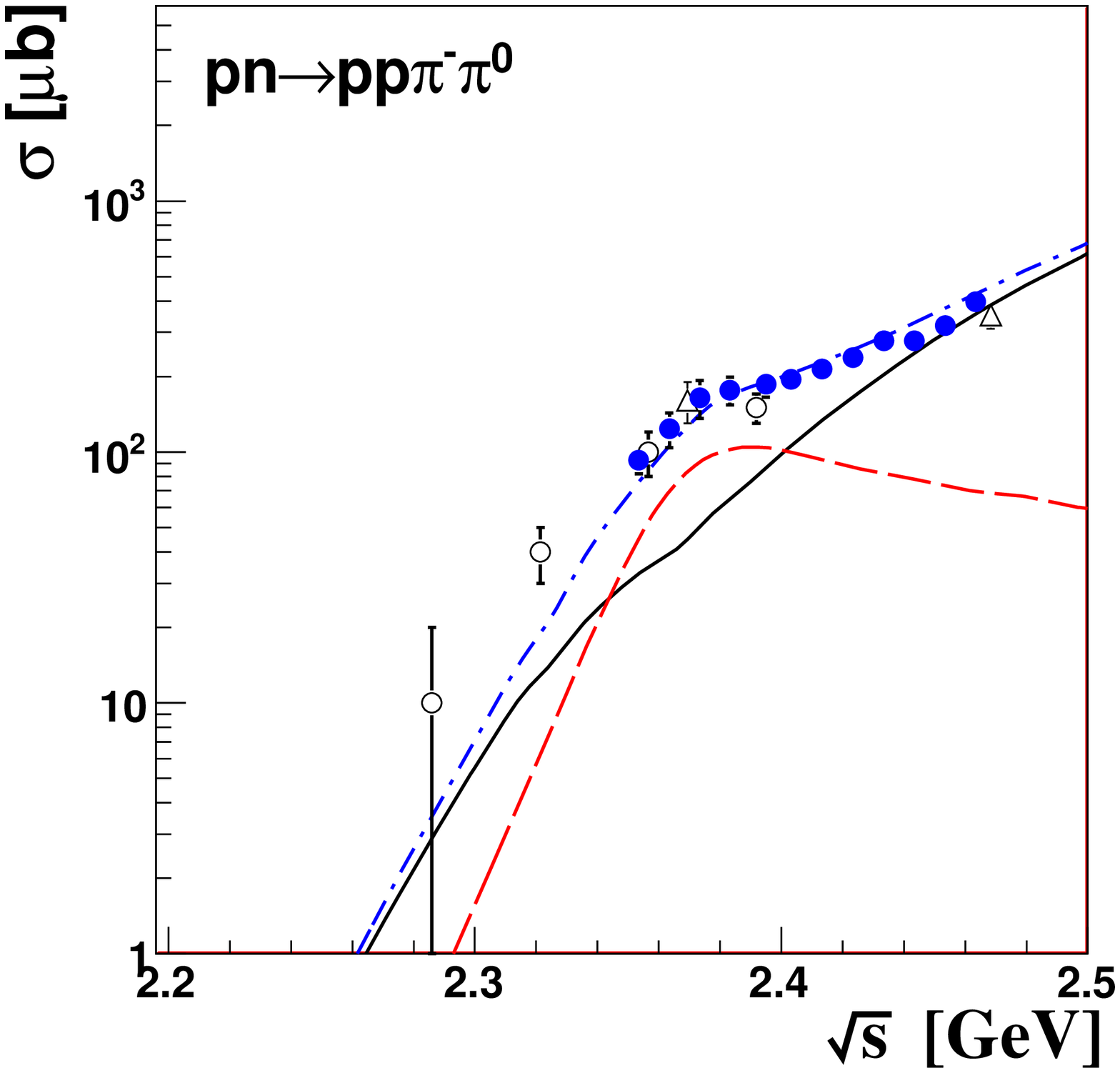}
\includegraphics[width=4.9cm,clip]{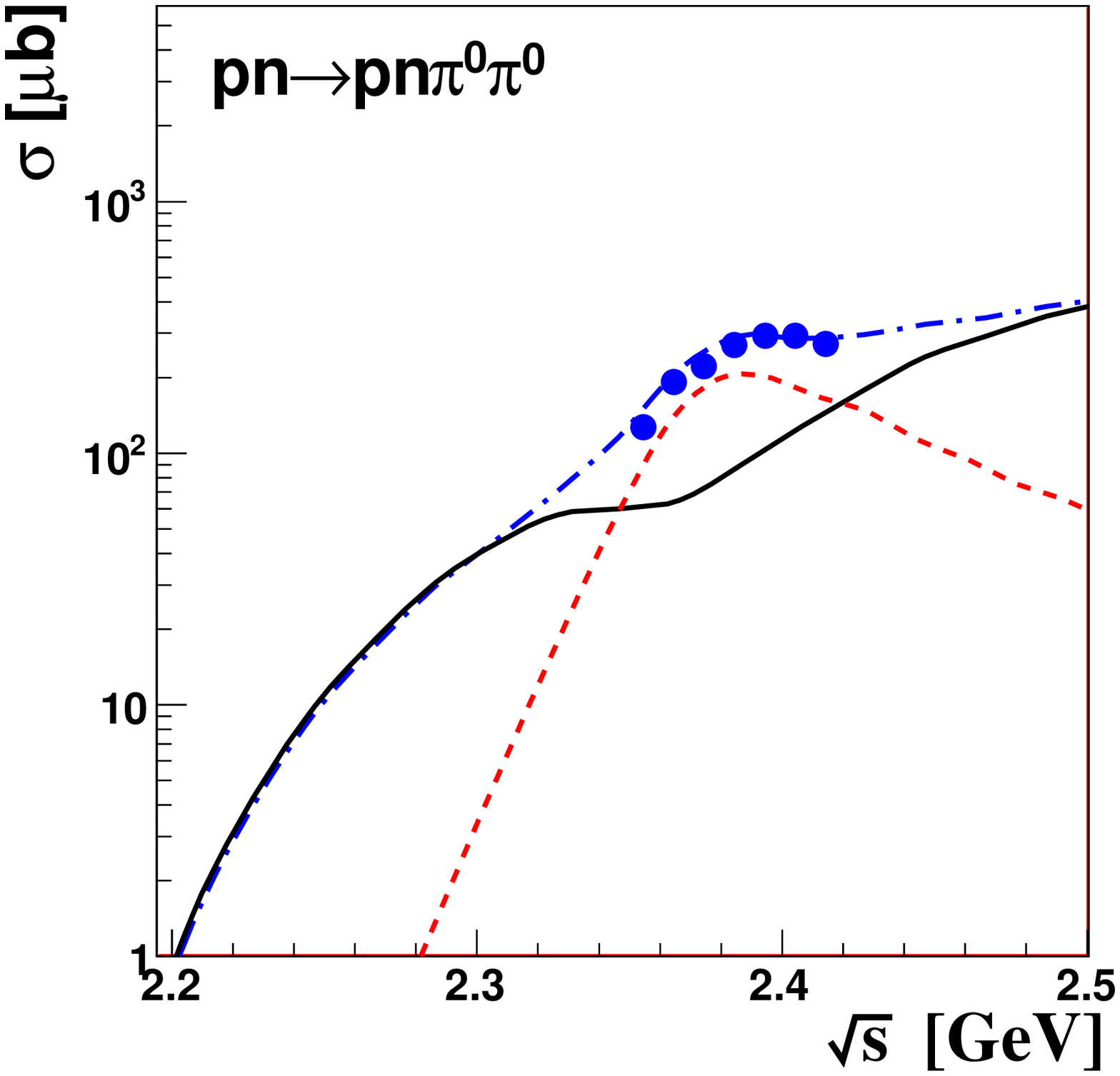}
\includegraphics[width=4.9cm,clip]{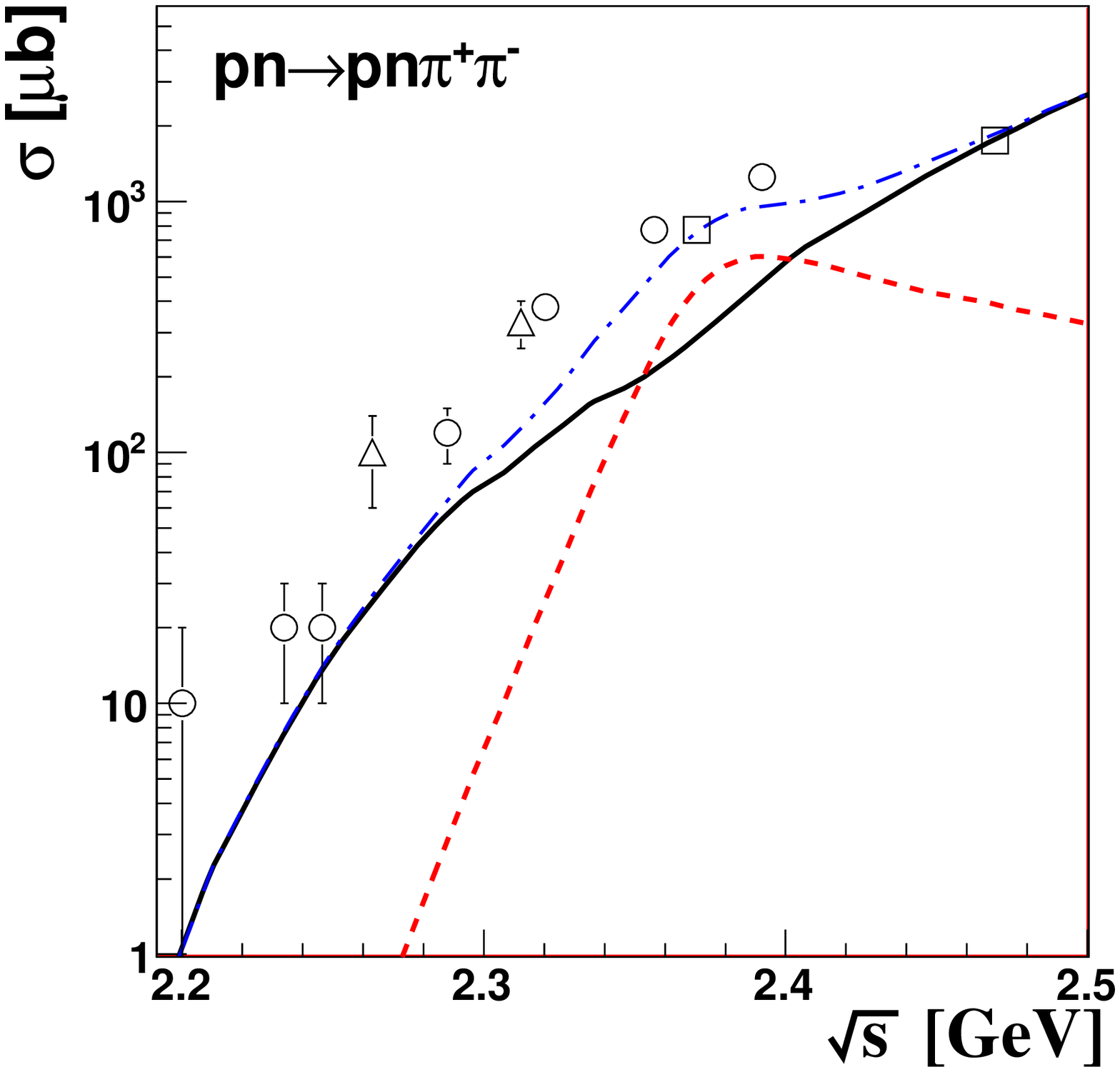}
\caption{Dependence of two-pion production to the non-fusion channels $pp
  \to pp\pi^-\pi^0$ (left), $pn \to pn\pi^0\pi^0$ (middle) and $pn \to
  pp\pi^+\pi^-$ (right) on the center-of-mass energy $\sqrt s$. Black, solid
  lines denote calculations of conventional $t$-channel processes 
 leading to Roper, $\Delta\Delta$ and $\Delta(1600)$ excitations 
in the modified Valencia model. Red, dashed lines show the $d^*$
contributions and blue, dash-dotted lines their coherent sum. Filled blue
circles give the WASA data, open symbols previous experimental results, see 
Refs. \cite{pp0-,np00,KEK,brunt,dubna}.}
\label{fig-2}       
\end{figure}

In addition to its decay into fusion channels this resonance should decay also
into all isospin allowed 
non-fusion two-pion production channels, {\it i.e.} the channels
$pp\pi^0\pi^-$, $pn\pi^0\pi^0$ and $pn\pi^+\pi^-$. Again, for lack of
appropriate data, the first two channels were studied with WASA at COSY in 
the energy region of interest. The results (blue solid dots) for the total
cross sections are shown in Fig.~2 together with data from previous
experiments (open symbols) \cite{pp0-,np00}. The experiments
are again consistent with the appearance of the $d^*$ resonance as expected
from isospin relations. Note that the resonance effect appears here not as
pronounced, since it sits on the steep slope of a strongly rising
four-body phase-space from conventional processes.

For the $pn\pi^+\pi^-$ channel there exist as of yet solely
bubble-chamber data \cite{KEK,brunt,dubna}. Within their very limited
precision they agree reasonably 
well with our calculation for this channel, as depicted in Fig.~2,
right. This calculation uses the same conventional background description as
used for the other $NN\pi\pi$ channels. For the resonance strength it utilizes
the prediction of Albaladejo and Oset \cite{oset} and includes also the
small $\rho$ channel configuration, not taken into account in
Ref.~\cite{oset}, but observed in the $pp\pi^0\pi^-$ channel (Fig.~23, left).
Since the HADES collaboration has performed measurements for this
channel in the energy region of interest, high-quality data may soon get
available.  

\section{Comparison to theoretical predictions}

A dibaryon state with $I(J^P) = 0(3^+)$ having an asymptotic $\Delta\Delta$
configuration was already predicted by Dyson and Xuong \cite{Dyson} and even
the predicted mass was close to the one observed now. Later on Goldman {\it et
  al.} pointed 
out the unique features of such an state based on its particular symmetries
due to its quantum numbers and called it the 
"inevitable dibaryon" \cite{Goldman}.

Recent state-of-the-art three-body \cite{GG1,GG2} and quark-model
\cite{ping,Zhang,Huang} calculations obtain this resonance at about the right
mass and partly also reproduce its width.

\section{Conclusions}
\label{sec-4}

For the first time a genuine dibaryon resonance has been identified by
observing it in various decay channels -- in particular also in the
elastic $pn$ channel. This experimental success was only possible
by kinematically complete (even overdetermined) and exclusive measurements
using the best suitable equipment consisting of high-precision
storage-ring beams, pellet target and a hermetic detector
covering the full reaction phase-space.

\ack{
We acknowledge valuable discussions with J. Haidenbauer, C. Hanhart,
A. Kacharava, E. Oset, I. Strakovsky, C. Wilkin and R. Workman on this
issue. This work has been supported 
by BMBF, Research Center J\"ulich (COSY-FFE) and DFG (CL 214/3-1).
}
\\

\end{document}